# Spurious peaks arising from multiple scattering events involving cryostat walls in inelastic neutron scattering

**Lothar Pintschovius[a], Dmitry Reznik[b], Frank Weber[a]\*, Philippe Bourges[c], Dan Parshall[b], Ranjan Mittal[d], Samrath Lal Chaplot[d], Rolf Heid[a], Thomas Wolf[a], Daniel Lamago[a] and Jeffrey W. Lynn[e]**

[a]Institute of Solid State Physics, Karlsruhe Institute of Technology, PO 3640, Karlsruhe, D-76021, Germany
[b]Department of Physics, University of Colorado, Boulder, Colorado, CO 80309, USA
[c]Laboratoire Léon Brillouin, CEA Saclay, Gif-sur-Yvette, F-91191, France
[d]Solid State Physics Division, Bhabha Atomic Research Centre, Trombay, Mumbai, 400 085, India
[e]NIST Center for Neutron Research, National Institute of Standards and Technology, Gaithersburg, Maryland, 20899-6012, USA

Correspondence email: frank.weber@kit.edu

**Synopsis** A new type of artefact is described, which was observed in a wide variety of inelastic neutron scattering spectra, and which can be easily mistaken as a signature of real excitations.

**Abstract** Well defined peaks with energies of about 18 meV have been observed in a variety of inelastic neutron scattering experiments on single crystals as well as on powders using either the triple-axis or the time-of-flight technique. They can easily be mistaken for signatures of real excitations. We have found that they are due to multiple scattering events involving primarily walls of the sample environment. Hence, they are particularly troublesome in experiments using very small samples as have been used with recently developed high intensity neutron spectrometers. It will be discussed what needs to be done to reduce the unwanted scattering to a minimum.

## 1. Introduction

In the course of a neutron scattering study of the phonons in $CaFe_2As_2$ broad, but well defined peaks were observed with energies of about 17 meV in a certain Brillouin zone (Mittal *et al.*, 2009) (the data look very similar to those shown in Fig. 1). These peaks could be followed from the zone boundary to the zone center, although with decreasing intensity. Their intensities, energies and dispersion were typical of optical phonons and therefore, they were assigned to a phonon branch expected to show up in this particular Brillouin zone. There was a moderate disagreement as to the energies of this phonon branch predicted by density functional theory, but this was not seen as a serious problem. The relatively large width of the peaks (about 6 meV full width half maximum) was taken as evidence of a strong coupling of these phonons to the electrons.

In a subsequent neutron scattering study of the phonons in slightly overdoped $BaFe_{1.8}Co_{0.2}As_2$ very similar peaks showed up in the same locations in reciprocal space, and hence were taken as genuine interesting features of 122-compounds (Pintschovius *et al.*, 2009). However, inelastic x-ray measurements on doped and undoped $BaFe_{2-x}Co_xAs_2$ did not show such peaks (Reznik *et al.*, 2009), which caused considerable concern about their origin. Since inelastic neutron scattering is known to be plagued by parasitic scattering, intensive checks were made to find out whether or not the 17 meV-peaks are instrumental artifacts. Using a variety of experimental set-ups, including time-of-flight spectrometers, all the major sources known to create accidental peaks could be safely excluded. In particular, it became clear that the 17 meV peaks are not caused by higher order problems of the monochromator or analyzer. On the other hand, the knowledge about the phonons in these systems had developed to such an extent that it could also definitely been ruled out that these peaks are associated with ordinary phonons.

We note that similar unexplained peaks had been observed in another investigation on pnictide samples. Mittal *et al.* (2008) and Zbiri *et al.* (2009) reported a peak at about 20 meV in the experimental data for the phonon density of states in $BaFe_2As_2$ for which they had no clear explanation. The inability of the theory to reproduce the 20 meV peak was seen as the main shortcoming of their density functional theory calculations.

Since these peaks were absent in spectra recorded by inelastic x-ray scattering, it was hypothesized that they are magnetic in origin. However, we found that that their intensity does not follow the magnetic form factor of iron (Fig.2 ). Additional measurements using polarized neutrons have definitely ruled out that the peaks are of magnetic origin. After an intense search for the origin of the unexplained peaks it became finally clear that they are indeed artifacts, but that they have a different origin from all the artifacts known so far, i.e. that they arise from double scattering processes involving walls of the sample environment. Hence, they are particularly troublesome in experiments using very small samples. In the following, it will be discussed what needs to be done to reduce the unwanted peaks to a minimum and to discriminate them against real excitations.

## 2. Experiments

A large number of measurements were performed on the 1T triple axis spectrometer at the Laboratoire Léon Brillouin (LLB), Saclay. Additional experiments were conducted on the BT7 triple axis spectrometer at NIST (Lynn *et al.*, 2012) and on the 2T triple axis spectrometer at LLB for the sake of polarization analysis. Moreover, we analyzed data which were taken using the time-of-flight (TOF) technique for a completely different purpose, i.e. data taken on powder samples using MERLIN at ISIS (Castellan *et al.,* 2011) and on a single crystal sample using the ARCS spectrometer at the SNS (Parshall *et al*., 2014). Further details will be given in the next Section when discussing specific results.

## 3. Results

The Results obtained at an early stage of our investigations are shown in Fig. 1. The sample had a composition of $Ba(Fe_{0.9}Co_{0.1})_2As_2$ . We note that the spectrum taken at Q=(2.5,1.5,0) looks very similar to that observed previously in $CaFe_2As_2$ (Mittal *et al.*,2009). Further, the decrease in intensity of the broad peak at about 18 meV when going from the zone boundary to the zone center was observed in $CaFe_2As_2$ as well. However, what we did not realize during the investigation of $CaFe_2As_2$, is the fact that the total intensity of all the peaks in the spectrum is larger than expected from the phonons. This fact comes out very clearly in Fig. 1 by comparing the red and blue lines. The blue lines were calculated from the predictions of DFT after having calibrated the inelastic scattering structure factors using measurements of long wavelength acoustic phonons, i.e. phonons for which the structure factors can be calculated directly from the structure. Obviously, DFT predicts the frequencies and intensities of individual phonons very well, but more importantly, it is very clear that the broad peak at 18 meV is an extra feature which cannot be explained by scattering from phonons.

This kind of analysis was triggered by the fact that no such peak at 18 meV was observed in inelastic x-ray scattering (IXS) measurements on doped and undoped $BaFe_2As_2$ (Reznik *et al.*, 2009). In the following, a variety of further pnictide samples were investigated by neutron scattering on a variety of instruments. An instructive result is plotted in the lower panel of Fig. 1. Although these data were obtained with a completely different kind of spectrometer than those plotted in the upper panel, they look surprisingly similar, which is very untypical for an instrumental artifact. The experiment performed on the ARCS TOF-spectrometer allowed us to determine the Q-dependence of the 18 meV over a wide range of momentum transfer (Fig. 2). We found that the intensity of the 18 meV neither follow the $Q^2$-dependence of phonons nor the magnetic form factor. This result alone argued against a magnetic origin of the 18 meV peaks. Additional experiments using polarization analysis confirmed that they are due to nuclear scattering.

Further data taken on a variety of samples invariably showed unexplained peaks at about 18 meV. However, their intensity varied from sample to sample indicating that they are an extrinsic feature (Fig. 3). We suspected that they might be caused by sample imperfections, but we had to assume an unreasonably high content of impurities to explain the quite high intensity of the 18 meV peaks in many of the experiments. Hydrogen impurities could not be ruled out in this way because of the exceptionally large neutron scattering cross section of hydrogen. However, hydrogen impurities could be could be finally ruled out as well as source of the 18 meV. In particular, our experiment using polarization analysis spoke against it, because hydrogen scattering is nearly completely spin-incoherent, whereas the 18 meV peaks were found to come from nuclear coherent scattering.

After an intense search for the reasons of the 18 meV peaks, the answer came from an experiment using a top-loading closed-cycle refrigerator (CCR), which allowed us to study the scattering with and without sample under otherwise identical conditions. These experiments revealed that scattering from cryostat walls was sufficient to generate such peaks (Fig. 3). Since slits in the incoming and the scattered beam prevented any neutrons scattered just once from reaching the detector, the recorded neutrons have to be scattered at least twice.

Apparently, subtracting empty cryostat scans allows one to largely eliminate the parasitic peaks to a very large extent. Remnants of it after the subtraction are probably due to the fact that scattering from the sample itself (and the material needed to attach it to the sample holder) is involved as well to some extent. Empty-can scans are rarely performed in experiments using triple-axis spectrometers, and would be, due to the Q and T dependencies of the parasitic scattering, very time consuming (an example demonstrating the T dependence is shown in Fig. 4). Therefore, we made a systematic investigation to find out whether a careful masking of the incoming and the scattered beam might be enough to largely eliminate the parasitic peaks. Unfortunately, the result of this study was negative: closing the slits in the incoming and the scattered beam progressively down, and repeating the same energy scan again and again, did understandably show a concurrent reduction of the parasitic scattering. However, there was still a sizeable parasitic peak when any further narrowing of the slits reduced the real signal as well (Fig. 5).

We note that optimizing the slits is a rather time-consuming procedure, and closing the slits to a point when Bragg intensities start to decrease will invariably lead to a significant loss in inelastic signal. This loss will depend, of course, on how large the divergencies of a particular instrument really are. We note that the divergencies used of the 1T-instrument at LLB, on which many of the results shown here were obtained, are on the low side of the ones used with actual triple-axis instruments around the world.

Masking the beam will be more efficient the closer the slits can be pushed towards the sample. Obviously, the dimensions of the equipment used as sample environment sets a lower limit. However, the motorized slits used on modern instruments can often not be put close to the cryostat walls because they are too bulky. In our case, additional slits were prepared which could be placed very close to the cryostat walls of diameter 20 cm. These slits did help, although a clear peak from parasitic scattering was still present (Fig.6).

From what was discussed above, it follows that empty-can scans are to our present knowledge the most resounding way to eliminate the parasitic 18 meV-peaks. Empty-can scans are routinely made in TOF-experiments using powder samples. Therefore, one might think that the parasitic scattering does not pose a problem in such experiments. Unfortunately, this is not fully true. An example is shown in Fig.7. As can be seen in that figure, the raw data show indeed a peak at about 20 meV which was not expected from the phonon properties of this particular sample (again a pnictide sample). After subtracting the empty-can scan, which shows a clear peak at this energy as well, the difference data still show too much intensity around 20 meV compared to the prediction of DFT. According to our understanding, this is due to the fact that the double scattering processes leading to the enhanced scattering around 20 meV may involve not only the sample environment, but the sample itself as well. We surmise that the enigmatic peaks observed at about 20 meV in neutron TOF-data on pnictide samples reported by Mittal *et al.* (2008) and Zbiri *et al*. (2009) came from the same type of parasitic scattering.

## 4. Discussion

Inelastic neutron scattering is notoriously plagued by artifacts. Several of them are well known and largely understood, at least in principle. Many checks were made to exclude the well-known artifacts due to higher orders in the incoming or scattered beam. The fact that the parasitic scattering described here shows a pronounced temperature dependence in that it follows approximately the Bose statistics and that the peak position decreases on heating by a couple of meV, misled us for a longtime to believe that the unexplained peaks are related to genuine excitations. The question arises why this particular artifact remained unknown in the neutron scattering community. One possible reason is that as far as triple-axis experiments are concerned, utilizing an $E_f$ of » 14 meV with a PG filter is a standard configuration, which has a known spurious eleastic scattering process from second order in the incident beam and third order in the scattered beam. Therefore unexpected scattering near this energy might already be discarded as spurious. Another important factor originates form the development of modern instrumentation that utilizes horizontal focusing. In the early days spectrometers were generally equipped with Soller collimators and flat monochromators, while now doubly-focusing monochromators and analyzers are available, opening up the instrumental resolution, with radial collimation or no collimators at all. This change has enhanced the intensity of the signals from vibrational or magnetic excitations but also from the sample environment as well, and with the smaller samples the contribution from the sample environment may be relatively larger. The ensuing problems were further aggravated by the fact that the large horizontal and vertical divergencies going along with doubly-focusing monochromators make it quite difficult to mask everything except the sample itself. In that sense, the troublesome artifact discussed here was the price to pay for the possibility to study samples much smaller than those necessary in the early days.

The parasitic scattering could not have been confounded with scattering from real excitations if it would not be peaked at a certain energy, in general between 17 and 20 meV. This energy seems to be connected to the maximum in the phonon-density of states (PDOS) of Al, although the energy distribution of the parasitic scattering is not a close replica of the PDOS of Al. Al is the most commonly used material for the direct sample environment because of its high transparency for neutrons. Using different materials for the sample environment will probably shift the peak of the parasitic scattering to other energies, but enhance rather than diminish its intensity, because of the higher neutron scattering power of other principally suitable materials, e.g. steel. Another option is, at least in principle, to keep all walls fairly distant from the sample. We found that using a container of large diameter (25 cm) and masking the direct sample holder with Cd does bring down the scattering to low levels. However, such a choice may not be practical for cryogenic equipment without some significant design changes.

As far as TOF-experiments are concerned, the best remedy to keep the parasitic scattering in check seems to be to keep the amount of material used for the sample environment to a strict minimum

## 5. Recommandations and Conclusions

We found that parasitic scattering from the sample environment gives rise to peaks at energies 17...20 meV. These peaks can be easily mistaken as signatures of real excitations, since they are observed with a variety of experimental set-ups, and show a marked temperature dependence. Depending on the sample size and the thickness and diameter of the walls of the sample environment, the strength of the peaks varies from insignificant to very troublesome. Careful masking of the incident and the scattered beams is important to reduce the intensity of the 18 meV peaks to a minimum, but it is insufficient to avoid them altogether. Empty-can scans are recommended in cases when unexpected peaks are observed with energies of about 18 meV. Empty-can scans are routinely performed in TOF experiments on powder samples. However, subtracting the empty-can data from the raw data will not completely eliminate the parasitic peaks, because they arise from multiple scattering events, and scattering from the sample itself may be involved as well in generating these peaks.

In more general terms, sample mounting, masking, and appropriately shielded sample environment equipment are always critical items in planning and running a clean neutron experiment, but are especially important when open collimation conditions are employed.  When using small samples, it is absolutely critical to use as small a sample mount as possible, to reduce the scattering from the mount with respect to the sample.  This can also help avoid scattering from exchange gas used for thermal equilibrium, especially at lower temperatures where it can condense on the sample and mount and give unwanted, temperature-dependent signals (Chi *et al.,*2011).  Glues should be avoided if possible; ones containing hydrogen must be avoided, but it should be noted that hydrogen-free glues scatter and can scatter quite strongly compared to the sample.  Glues can also cause substantial strain to the sample as the temperature is varied, which can change the physical properties substantially.  Finally, sample environment equipment should be shielded inside as much as possible—a consideration primarly for sample-environment teams—especially above and below the beam line, as the spurious process considered here involves multiple scattering events.  To get the highest quality data possible, it is in the best interests of the experimenter to insure that the sample holder is small and masked as much as possible, that the area around the sample absorbs neutrons in all areas other than the incident and scattered beams, and that those beams are reduced in size to just illuminate the sample from both the monochromator and detector(s) sides.

**Acknowledgements**    We thank R. Osborn and S. Rosenkranz for helpful discussions and making us their raw data (of Ref. 6) available. D.P and D.R. were supported by the DOE, Office of 329 Basic Energy Sciences, Office of Science, under Contract 330 No. DE-SC0006939. The research at ORNL's Spallation 331 NeutronSource was sponsored by the Scientific User Facilities 332 Division, Office of Basic Energy Sciences, U.S. Department 333 of Energy.

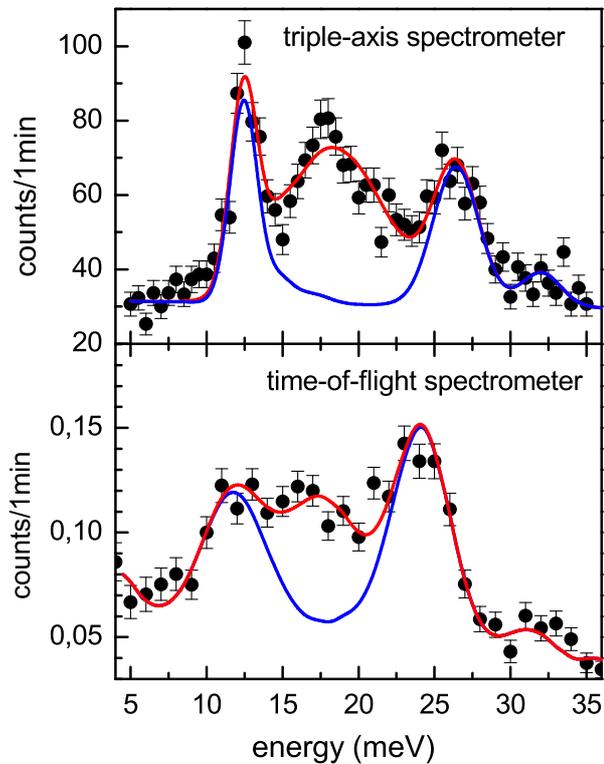

**Figure 1** Energy scans taken on pnictide samples at T=11 K using two different kinds of spectrometers, i.e. a triple-axis spectrometer (spectrometer 1T at LLB, Saclay) and a time-of-flight spectrometer (ARCS at the SNS, Oak Ridge). The momentum transfer was Q=(2.5,1.5,0) in both cases. The two samples were of somewhat different composition (above: $BaFe_{1.9}Co_{0.1}As_2$, below: $SrFe_2As_2$), but the phonon peaks expected at this momentum transfer are very similar. The blue lines correspond to the phonon intensities expected from DFT plus a linear background. The red lines are fits to all peaks in the spectra. Error bars are statistical in origin and represent one standard deviation unless otherwise noted.

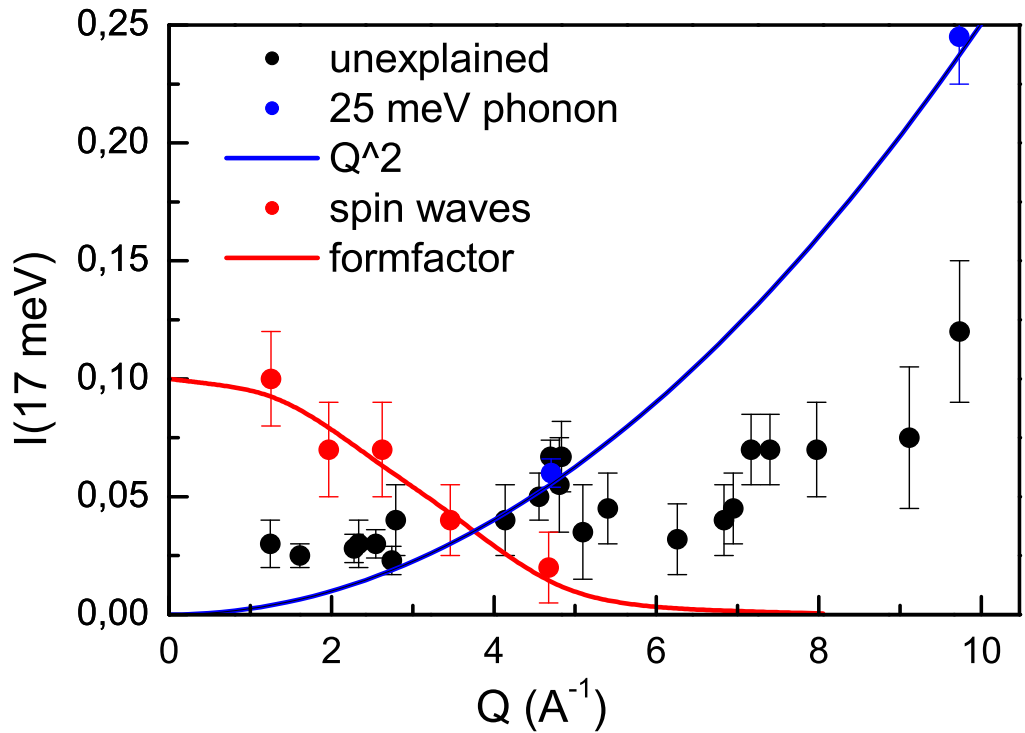

**Figure 2** Q-dependence of the 18meV-peaks evaluated from TOF-data taken on an undoped Sr122-sample at T = 10 K. These data are a byproduct of an experiment with a different purpose (Parshall *et al.*, 2014). The error bars reflect not so much statistical errors rather than uncertainties from incomplete knowledge of the background and of the contribution of normal phonons. The spin wave intensities were evaluated at 18 meV as well. Note that the intensity ratio of peaks related to conventional spin waves and to the unexplained excitations depends on the binning of the data with respect to the resolution of the instrument, since the spin wave intensities are sharply peaked in momentum space, whereas the others are not. The solid red line depicts the Fe form factor. Data points in blue correspond to intensities of ordinary phonons with energy of 25 meV, taken from scans with about the same inelastic structure factor. Since the phonon peaks are resolution limited, whereas the 18 meV peaks are two times as broad, the phonon intensities were divided by a factor 2.

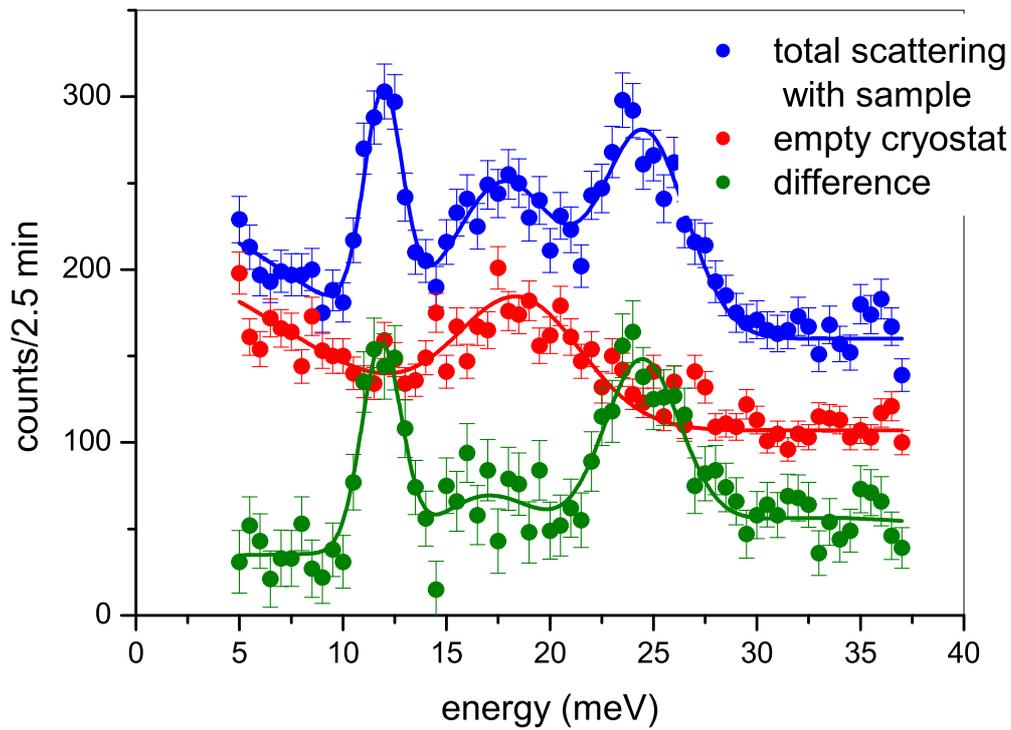

**Figure 3** Blue dots: Energy scan taken on a small sample of optimally doped Ba-122 at room temperature. The momentum transfer was the same as that in Fig. 1. The sample was wrapped in a piece of thin aluminum foil, attached to the bottom of a stick of a top-loading cryostat and inserted into the sample chamber of 5 cm diameter. The blue line is the result of a fit with three Gaussians plus three parameters describing the background. Red dots: Energy scan after raising the stick to an extent that the sample was completely out of the beam. Olive dots: Difference of the two above scans. The data points and the fit curve have been down-shifted by 50 counts for the sake of clarity. Note that the raw data have been up-shifted by 50 counts for the sake of clarity.

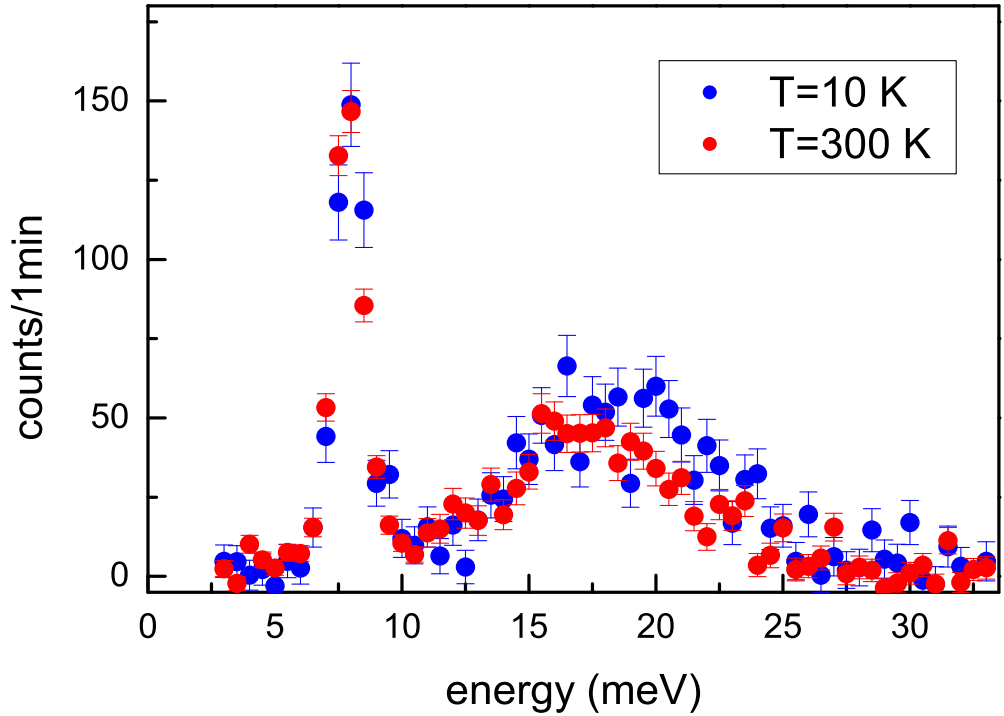

**Figure 4** Energy spectra taken on BaFe$_2$As$_2$ at two different temperatures. A linear background was subtracted in both cases. In addition, the 300 K-data have been corrected for the Bose-factor of the peaks observed at E=7.4 meV (factor of 3.9) or 18 meV (factor of 1.95), respectively. The peak at E=7.4 meV is associated with a transverse acoustic phonon. All other phonons have a very low inelastic structure factor at this particular wave vector, i.e. Q=(2,2,1).

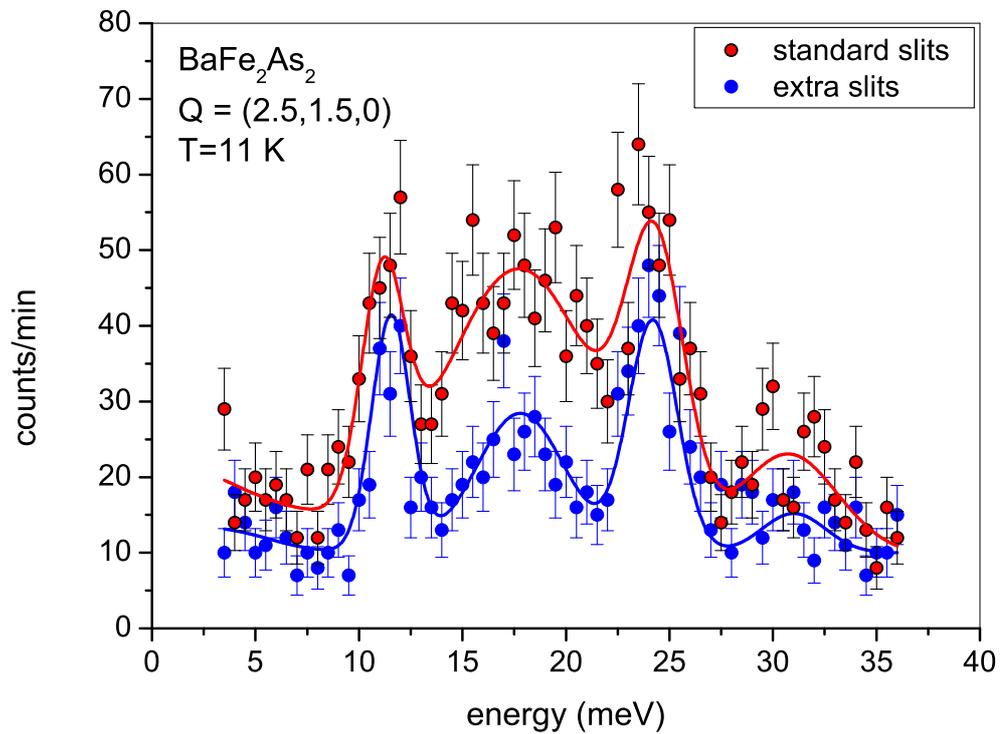

**Figure 5** The two energy scans shown above were taken on the same sample on the same spectrometer with the same settings. The only difference was the insertion of extra slits made of $B_4C$-filled plastics at positions close to the surface of the cryostat, whereas the standard slits had to be kept more than 10 cm further away because of their bulkiness. The sample was a platelet of $BaFe_2As_2$ with mass of about 1 g. The particular Q was chosen for the sake of comparison with many scans performed previously.

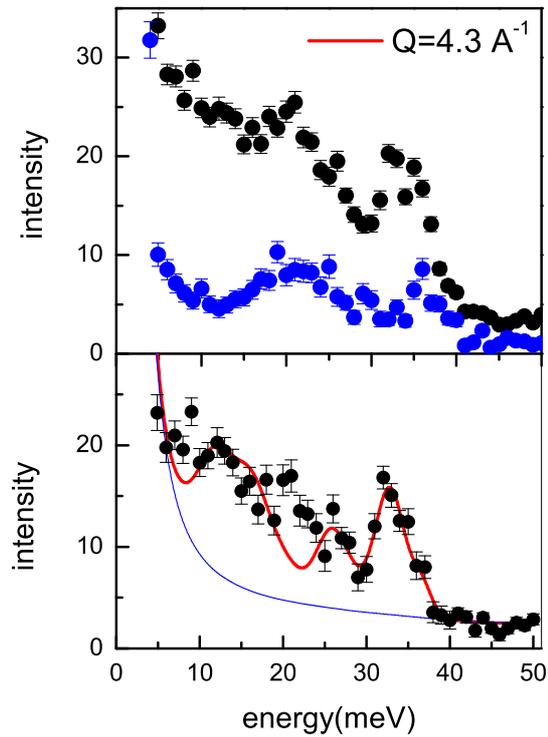

**Figure 6** Upper panel: neutron intensity observed on a powder sample of $Ba_{0.6}K_{0.4}Fe_2As_2$ by the time-of-flight technique at T=7 K in the range 4.2 $A^{-1}$<Q<4.5 $A^{-1}$ (black dots). The data were taken on the MERLIN spectrometer at ISIS, Rutherford laboratory, using an incident energy of 60 meV. Data taken with an empty can are shown by blue dots (the data are a byproduct of the experiment published by Castellan *et al.* (2011)). Lower panel: the difference between the raw data and the empty-can data are shown as black dots. The experimental result was simulated on the basis of DFT-results for the Q-dependent neutron intensities, the instrumental resolution plus an estimate for the background as shown by the blue line. The calculated results (red line) were normalized in a way as to approximately describe the observed intensities around 33 meV.

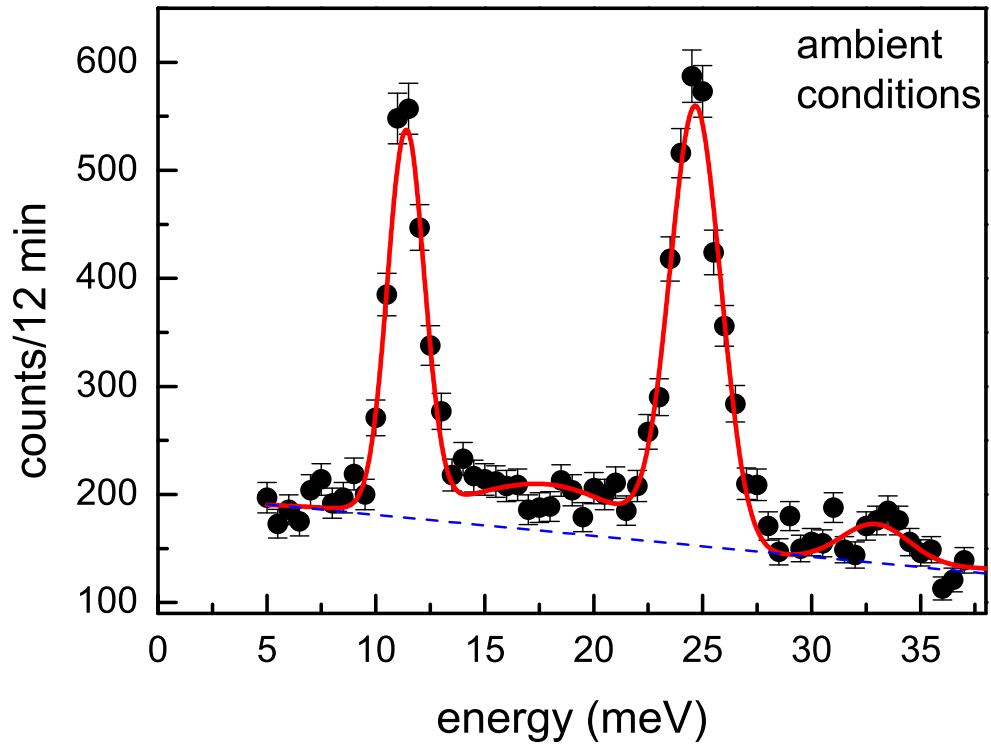

**Figure 7** Energy scan taken on a relatively small pnictide sample (~1 g) at ambient conditions. The sample was mounted on a platform made of Cd and placed at the center of a hollow aluminium cylinder (diameter 25 cm) in vacuum. The momentum transfer was the same as that used in Figs. 1,4,5, at which a strong peak at E=18 meV had been invariably observed using standard cryogenic equipment.